\documentclass[useAMS,usenatbib]{mn2e}

\newcommand{\aap}{A\&A}
\newcommand{\apj}{ApJ}
\newcommand{\apjs}{ApJS}

\newcommand{\araa}{ARA\&A}
\newcommand{\nat}{Nature}
\newcommand{\mnras}{MNRAS}
\newcommand{\lsun}{\ensuremath{\mbox{L}_{\odot}}}
\newcommand{\ghz}{\ensuremath{\mbox{GHz}}}

\newcommand{\kms}{\ensuremath{\mbox{km\,s}^{-1}}}
\newcommand{\mpc}{\ensuremath{\mbox{Mpc}}}

\newcommand{\snu}{\ensuremath{S_{\nu}}}
\newcommand{\lnu}{\ensuremath{L_{\nu'}}}
\newcommand{\hsf}{\ensuremath{h_{\rm 71}}}
\newcommand{\lfir}{\ensuremath{L_{\rm FIR}}}
\newcommand{\loh}{\ensuremath{L_{\rm OH}}}
\newcommand{\lrad}{\ensuremath{L_{\rm radio}}}

\input{psfig.sty}

\def\gs{\mathrel{\raise0.35ex\hbox{$\scriptstyle >$}\kern-0.6em
\lower0.40ex\hbox{{$\scriptstyle \sim$}}}}
\def\ls{\mathrel{\raise0.35ex\hbox{$\scriptstyle <$}\kern-0.6em
\lower0.40ex\hbox{{$\scriptstyle \sim$}}}}
\def\m@th{\mathsurround=0pt }
\def\eqalign#1{\null\,\vcenter{\openup1\jot \m@th
 \ialign{\strut\hfil$\displaystyle{##}$&$\displaystyle{{}##}$\hfil
 \crcr#1\crcr}}\,}

\title[Searching for a gigamaser in APM\,08279+5255]
      {Searching for a gigamaser in APM\,08279+5255, and other short stories}
\author[Ivison]
       {R.\,J.\ Ivison$^{1,2}$
        \vspace*{1mm}\\
        $^1$ UK Astronomy Technology Centre, Royal Observatory, Blackford Hill,
             Edinburgh EH9 3HJ\\
        $^2$ Institute for Astronomy, University of Edinburgh, Blackford Hill,
             Edinburgh EH9 3HJ
}

\date{Accepted 2006 April 28. Received 2006 April 26; in original form 2006 March
28}

\pagerange{000--000}

\begin{document}

\maketitle

\begin{abstract}
Bolometer arrays on large antennas at high, dry sites have unveiled a
dusty population of massive, luminous galaxies -- submillimetre
galaxies, or SMGs -- which make a significant contribution to the
star-formation rate density at $z>\rm 1$. The most crucial piece of
information required to derive the history of obscured star formation
is the redshift distribution of this galaxy population, $N(z)$, which
breaks degeneracies in the models and allows the mass and dynamics of
the galaxies to be explored via high-resolution three-dimensional
imaging in CO and by determining their level of clustering. Many SMGs
are extremely faint, optically; some have no plausible counterparts,
even in the infrared (IR), making the determination of an unbiased
$N(z)$ very difficult. The arrival of the {\em Herschel Space
Observatory} and next-generation ground-based submm cameras will
likely exacerbate this so-called `redshift deadlock'. Here, we report
the first test of a new method for determining redshifts, based on the
observed dependence of maser and IR luminosities. We have searched the
dusty, lensed, hyperluminous quasar, APM\,08279+5255, for the 1612-,
1665- and 1667-MHz hydroxyl lines as well as the 22-GHz water line. At
$z$ = 3.9 these are shifted to 329, 340 and 4,538\,MHz. Our relatively
shallow test data reveal no convincing maser activity but we set a
meaningful constraint on the OH maser luminosity and we approach the
expected thermal noise levels, meaning progress is possible. As an
aside, we present deep new submm and radio imaging of this
field. Using a simple shift-and-add technique we uncover a new submm
galaxy, conceivably at the redshift of APM\,08279+5255.
\end{abstract}


\section{Introduction}

The SMG population, discovered using the SCUBA bolometer array
(Holland et al.\ 1999), has led astronomers a merry dance for almost a
decade (Blain et al.\ 2002).  Early indications, based on the SCUBA
Lens Survey (Smail et al.\ 2002), painted SMGs as an extraordinarily
diverse population, some red, some blue, some starbursts, some with
unequivocal AGN features (Ivison et al.\ 1998, 2000). However, over
the years some common characteristics have become familiar: on the
whole, SMGs are luminous ($\lfir\gs\rm 10^{12-13} \hsf^{-2} \lsun$,
where $\hsf \equiv H_{0}/\rm 71\,\kms \mpc^{-1}$) optically faint,
red, morphologically disturbed, massive and distant (Smail et al.\
1999; Gear et al.\ 2000; Chapman et al.\ 2003; Webb et al.\ 2003; Neri
et al.\ 2003; Blain et al.\ 2004; Pope et al.\ 2005; Greve et al.\
2005).

Due in large part to the spectroscopic survey of Chapman et al.\
(2005) we have seen a good deal of recent progress towards determining
one of the most crucial parameters in galaxy-formation models: the
redshift distribution of SMGs, $N(z)$. Knowledge of $N(z)$ breaks
degeneracies in the models (e.g.\ Blain et al.\ 1999); it also
provides an estimate of their typical mass via the clustering
characteristics of the population and, for individual galaxies, via
observations of CO, C\,{\sc i}, etc. (Frayer et al.\ 1998, 1999; Neri
et al.\ 2003; Greve et al.\ 2005; Papadopoulos 2006).

However, the available spectroscopic redshift distribution for SMGs is
biased: a radio detection is required to provide positional
information (e.g.\ Ivison et al.\ 2002), and rest-frame ultraviolet
emission lines are usually required to signal redshifts. It is not
clear that all SMGs will display these traits, certainly not the
dustiest systems at the highest redshifts (e.g.\ Berger et al.\
2006). Furthermore, it is not clear how redshifts will be determined
for the many thousands of dusty galaxies expected to be detected in
wide-field surveys planned for SCUBA-2 (Audley et al.\ 2004) and the
{\em Herschel Space Observatory} (Pilbratt 2004).

Powerful OH masing is relatively common amongst IR-luminous galaxies:
the most recent survey (Darling \& Giovanelli 2002) found that at
least a third of ultraluminous IR galaxies (ULIRGs, $\lfir\ge
10^{12}\lsun$) support megamasers or gigamasers ($\loh \ge 10 \lsun$
or $\loh \ge 10^{3} \lsun$, respectively). If starbursts are
responsible for a significant fraction of the luminosity of
ultraluminous and hyperluminous IR galaxies, as currently thought
(e.g.\ Farrah et al.\ 2002), then the associated turbulence may enable
low-gain unsaturated masing (Burdyuzha \& Komberg 1990). The earliest
OH maser-line observations appeared to demonstrate a quadratic
relationship between OH and far-IR (FIR) luminosities, $\loh$ and
$\lfir$. This is believed to be due to the abundant flux of FIR
photons pumping an OH population inversion in the star-forming
molecular gas (Baan 1985, 1989). Since FIR and radio luminosities
($\lrad$) are well correlated (e.g.\ Helou, Soifer \& Rowan-Robinson
1986), emission stimulated by the background radio continuum would
then yield $\loh\propto\lfir\lrad\propto\lfir^2$.

Townsend et al.\ (2001) argued that this quadratic dependence would
yield OH masers detectable amongst the SMG population, with peak
luminosity densities up to two orders of magnitude greater than those
seen in current samples. This would allow us to determine an accurate
and relatively unbiased $N(z)$ for SMGs, as well as constraining the
mass of their black holes, determining their geometric distances, even
probing the evolution of fundamental constants (e.g.\ Barvainis \&
Antonucci 2005; Lo 2005; Kanekar et al.\ 2005; Caproni, Abraham \&
Mosquera Cuesta 2006). Correcting large maser samples for Malmquist
bias (Kandalian 1996) favours a weaker dependence of $\loh$ on $\lfir$
than originally thought, $\loh \propto \lfir^{1.2\pm 0.1}$ (Darling \&
Giovanelli 2002); on the other hand, using a smaller, complete sample
continues to suggest a quadratic relationship, $\loh \propto
\lfir^{2.3\pm 0.6}$ (Kl\"ockner 2004). Either way, it is reasonable to
expect OH maser emission from a high proportion of SMGs, with $\loh$
in the range 10$^{4-5} \lsun$ for $\lfir\sim\rm 10^{13} \lsun$ and
mJy-level peak flux densities.

Maser searches have clear advantages over other methods of determining
$N(z)$ for SMGs: i) the bandwidth requirement for blind detection of
OH megamasers is low, $<$1\,$\ghz$ ($\nu_{\rm obs}$ = 165--835\,MHz
for $z$ = 1--10), smaller still with additional redshift constraints,
and the technique avoids contamination by H\,{\sc i} emission from
local galaxies (Briggs 1998) since there is no significant SMG
population at $z<\rm 1$ (Chapman et al.\ 2005); ii) the instantaneous
survey area is limited only by the primary beam (several square
degrees for an OH line search with the Giant Metre-wave Radio
Telescope -- GMRT); iii) interferometry permits some rejection of
radio-frequency interference (RFI); iv) the position of an emission
line can be pinpointed accurately, tying an emission line to an SMG
unequivocably (to a few arcsec, at worst); finally, v) the dual-line
1665-/1667-MHz OH spectral signature can act as a check on the line
identification and the reality of detections.

Here, we report a search for such maser emission. Blind searches for
maser emission from SMGs will not be feasible until next-generation
correlators are available, with their large instantaneous
bandwidths. To begin testing this technique we must therefore choose a
well-studied, luminous FIR galaxy, one with its redshift accurately
determined (see also Kl\"ockner 2004).

APM\,08279+5255 (Irwin et al.\ 1998) is an obvious candidate: a dusty,
broad-absorption-line (BAL) quasar embedded in a gas-rich starburst
with a prodigious FIR luminosity ($\gs10^{14} \lsun$). HCN and several
CO rotational transitions have been detected (Downes et al.\ 1999;
Papadopoulos et al.\ 2001; Wagg et al.\ 2005) yielding an accurate
systemic redshift: $\rm 3.91126\pm 0.00016$ (Axel Weiss, private
communication). Finally, adding to its appeal, it is amplified
gravitationally by a factor $A\sim\rm 7$ (Lewis et al.\ 2002) to yield
an apparent luminosity of $\sim$5\,$\times$\,10$^{15} \lsun$. At $z$ =
3.9, its OH and H$_2$O maser lines are both accessible to receivers at
the National Radio Astronomy Observatory's Very Large Array
(VLA). Even adopting $\loh\propto \lfir^{1.2}$ and an extreme lensing
model ($A=\rm 1000$, for which we have no evidence) we would
expect $\loh\sim 3\times 10^3 \lsun$. For a line width of
150\,km\,s$^{-1}$ we would then predict a flux density of almost 50\,mJy
after amplification, within reach of current facilities.

We also report new submm, radio and IR imaging which led to the
detection and characterisation of a new SMG in the vicinity of
APM\,08279+5255.

\section{Searching for OH and H$_2$O maser emission}

\begin{figure}
\centerline{\psfig{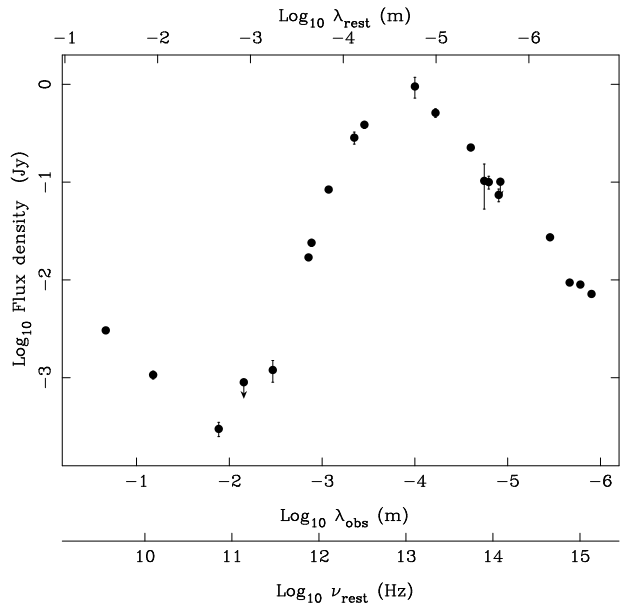}}
\vspace{-0.5cm}
\noindent{\small\addtolength{\baselineskip}{-3pt}}
\caption{The IR-to-radio SED of APM\,08279+5255, with radio data
from this paper and a compilation of data shortward of 5\,GHz (Lewis
et al.\ 1998; Downes et al.\ 1999; Egami et al.\ 2000; Papadopoulos et
al.\ 2001; Barvainis \& Ivison 2002; Soifer et al. 2004; Beelen
et al.\ 2006).}
\end{figure}

The choice of configuration for our VLA P-band observations
($\sim$340\,MHz) was dictated by the need to reduce confusion and
minimise RFI. We therefore opted for A configuration, obtaining
4.7\,hr of integration during 2002 April on the redshifted OH
lines. Approximately 80 per cent of the data were unaffected by RFI.

Data were taken with correlator mode 4, with one IF pair centred on
the 1612.231-MHz hyperfine OH line and another IF pair covering the
1665.4018- and 1667.359-MHz lines. Each IF had 61 $\times$ 48.83-kHz
channels, yielding a velocity coverage of
2,500\,km\,s$^{-1}$. Calibration scans of 0834+555, 0542+498 and
3C\,274 (1230+123) were obtained, to calibrate phase, amplitude and
absolute flux density, and to determine bandpass corrections.

The APM\,08279+5255 field is relatively devoid of bright radio
sources, so we were able to approach the thermal noise, with an
r.m.s.\ level of 0.5\,mJy\,beam$^{-1}$ in a pseudo-continuum image
made using all the spectral-line data, and as little as
3.4\,mJy\,beam$^{-1}$ in individual channel maps. Indeed,
APM\,08279+5255 is clearly detected in the pseudo-continuum P-band
map, with an integrated 334-MHz flux density of $\rm 5.1\pm 0.8$\,mJy.
It is not clear why, but one IF pair produced 2$\times$ deeper images
than the other.

We also obtained approximately 10\,hr of data with the VLA in its C
configuration during a flexibly scheduled block of time in 2001
August, using the C-band (5-GHz) receivers centred on the redshifted
22,235.08-MHz $\rm 6_{16}-5_{23}$ H$_2$O line. Correlator mode 2AD was
used: a single IF pair with 31 $\times$ 390.63-kHz channels giving
760\,km\,s$^{-1}$ of velocity coverage.  3C\,48 (0137+331) was used to
set the flux density scale and 0834+555 (5.7\,Jy at 4,527\,MHz) was
once again used as the local amplitude/phase calibrator. Again,
APM\,08279+5255 is detected in the pseudo-continuum map with an
integrated flux density of $\rm 1.07\pm 0.10$\,mJy. Fig.~1 shows the
overall radio spectral energy distribution (SED) of APM\,08279+5255,
adequately characterised by a power law of the form $S_{\nu}\propto
\nu^{-0.6}$, together with submm and IR data from the literature.

For both the C (H$_2$O) and P (OH) spectral-line data, image cubes
were made using AIPS {\sc imagr}, with {\sc clean} boxes placed
according to the positions of bright sources in the respective
pseudo-continuum maps. Cubes were made with 1- and 2-channel averages
and were inspected visually for signs of emission. Spectra were
extracted as follows: APM\,08279+5255 and three bright ($>$30-mJy),
compact, nearby sources were fitted ({\sc jmfit}) in the
pseudo-continuum images with single Gaussians, their sizes fixed to
that of the synthesised beam ($\sim$5.5\,arcsec {\sc fwhm}). Further
fits were then made in each spectral plane of each image cube, now
with both object sizes and pseudo-continuum positions fixed. The
best-fit peak flux densities were thereby extracted for each channel.
This technique delivers spectra with noise levels close to those
measured in individual image planes and remains relatively free of
other biases. The resulting spectra of APM\,08279+5255 were divided by
the respective data for the three bright, nearby sources, which had
been co-added and normalised by their total mean flux density. This
`flat field' was also used to provide a check on the reality of any
features in the spectra of APM\,08279+5255. Noise estimates in each
channel were obtained from source-free regions around the quasar using
{\sc imstat}. A simpler approach using {\sc possm} gave similar
results.

\begin{figure}
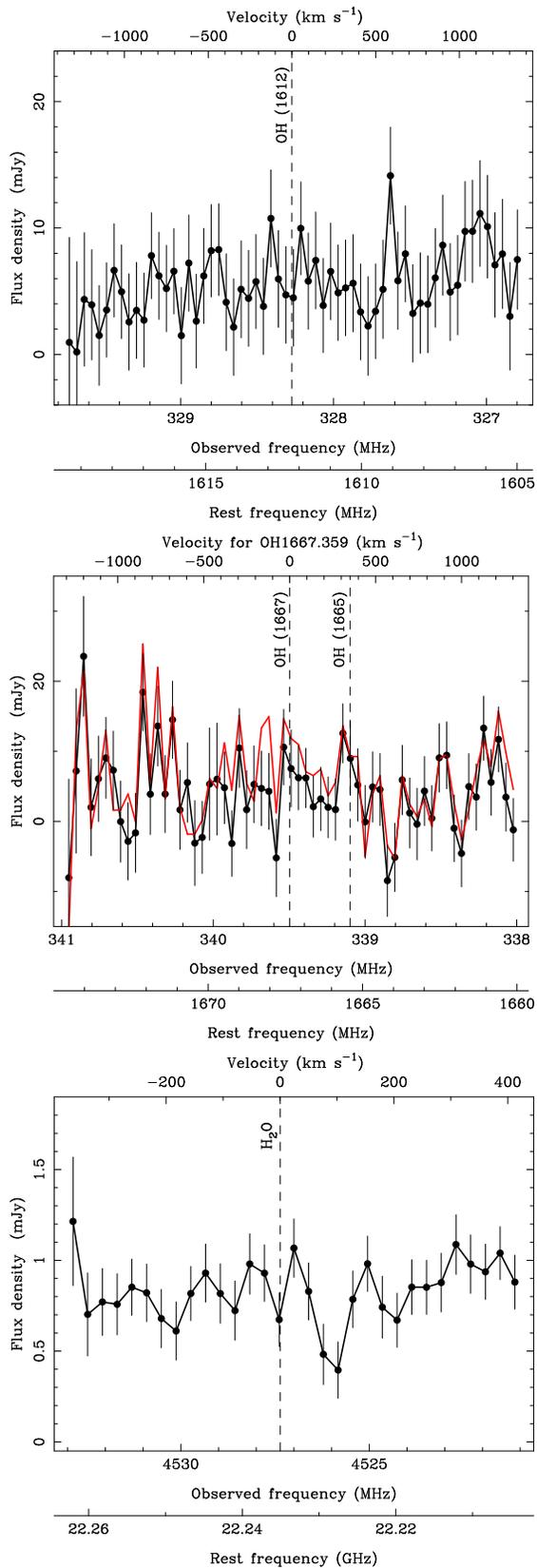

\centerline{\psfig{file=fig2a.eps,angle=270,width=2.9in}}
\vspace{0.1cm}
\centerline{\psfig{file=fig2b.eps,angle=270,width=2.9in}}
\vspace{0.1cm}
\centerline{\psfig{file=fig2c.eps,angle=270,width=2.9in}}
\vspace{-0.5cm}
\noindent{\small\addtolength{\baselineskip}{-3pt}}
\caption{Unsmoothed spectra covering the 1612-, 1665- and 1667-MHz OH
lines (top and middle) and the 22,235-MHz H$_2$O line
(lower). Uncertainties (1\,$\sigma$) are illustrated with error bars.
Line positions are marked and velocities are given for an assumed
redshift of 3.91126. Baselines have not been subtracted. The result of
extracting a spectrum with {\sc possm} are shown in red in the middle
panel.}
\end{figure}

\begin{figure}
\centerline{\psfig{file=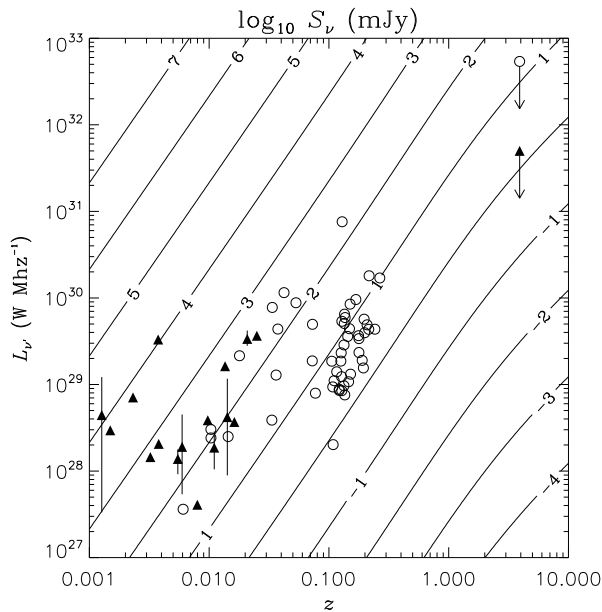,angle=0,width=3.3in}}
\vspace{-0.8cm}
\noindent{\small\addtolength{\baselineskip}{-3pt}}
\caption{Contour map of $\log_{10} \snu$, the peak observed flux
density of an OH maser, as a function of redshift, $z$, and peak
rest-frame luminosity density, $\lnu$, adapted from Townsend et al.\
(2001), for $\Omega_m=0.27, \Omega_\Lambda=0.73,
H_0=71$\,km\,s$^{-1}$\,Mpc$^{-1}$. H$_2$O and OH masers, in AGN and
IR-luminous galaxies, respectively, are shown as triangles and
circles, respectively.}
\end{figure}

The resulting spectra are shown in Fig.~2 where no convincing emission
from either OH or H$_2$O can be seen.  The two weak peaks seen at the
expected frequencies of OH (1665 and 1667\,MHz) are no more
significant than the dip seen in the H$_2$O spectrum at
$\sim$100\,km\,s$^{-1}$. The r.m.s.\ noise levels in the regions
around the OH lines at 1612 and 1665/1667\,MHz are 2.9 and 5.8\,mJy,
with 0.17\,mJy for the region around the H$_2$O line.  These figures
compare reasonably well with the theoretical expectations:
$\sim$2.9\,mJy in P band and $\sim$0.16\,mJy in C band, for good
weather and 26 available antennas.

There is a very weak positive feature at zero velocity in the
amplitude spectrum of the cross-correlation between the 1665-/1667-MHz
spectrum shown in Fig.~2 and a matched filter corresponding to two
Gaussian profiles with a separation fixed to that of the 1665- and
1667-MHz OH lines. We regard this as at best tentative evidence of
emission.

Given the very limited total bandwidth (and the poor resulting
velocity coverage) we cannot be sure that a broad line does not cover
the entire spectral window: the IRAS\,14070+0525 OH gigamaser ($z=\rm
0.265$, Baan et al.\ 1992) would certainly do so, though we would have
had no problem with a more typical linewidth ($\sim$150\,km\,s$^{-1}$,
Darling \& Giovanelli 2002). However, the peak flux density of such an
OH line cannot exceed our measured continuum levels, which are well
below the limits we would quote based on the OH spectral noise. For
the H$_2$O line, where this is not true, we add the continuum level to
the limit. The peak line flux densities from APM\,08279+5255 are thus
3$\sigma$ $<$ 8.7\,mJy for the satellite line of OH at 1612\,MHz,
17.4\,mJy for the main lines at 1665/1667\,MHz and 1.6\,mJy for
H$_2$O. For a rectangular line of width 150\,km\,s$^{-1}$ we have
constrained the lens-amplified OH luminosity to below $1.3\times 10^6
\lsun$ (3\,$\sigma$). As outlined in \S1, the apparent FIR luminosity
of APM\,08279+5255 exceeds that of IRAS\,14070+0525 by three orders of
magnitude so a detection of APM\,08279+5255 was quite conceivable and
the limit we have set for the main lines is significant, lying below
our most pessimistic prediction.  We note, however, that at least half
of local ULIRGs are not detected and that FIR luminosity is clearly
not the sole criterion for OH masing.

Following Townsend et al., the observed flux density $\snu$\ at a
frequency $\nu \equiv \nu'/(1+z)$ of a maser emitting with a peak
rest-frame isotropic luminosity density of $\lnu$ is given by $\snu =
(1+z) \lnu/(4\pi D_{L}^{2}(z))$, where $D_{L}(z)$ is the luminosity
distance. The $(1+z)$ factor in this expression accounts for the line
width narrowing due to the redshift, and partially offsets the
quadratic drop-off in $\snu$\ with distance. As an example, the peak
flux density of 10\,mJy for the IRAS\,14070+0525 OH gigamaser at
$z=\rm 0.265$ ($D_{L}(z)=\rm 4.1\times 10^{25}$\,m for $\Omega_m=0.27,
\Omega_\Lambda=0.73, H_0=71$\,km\,s$^{-1}$\,Mpc$^{-1}$), translates
into $\lnu=1.7\times 10^{30}$\,W\,MHz$^{-1}$.

Our limits on maser-line peak flux densities are therefore $\lnu<\rm
2.7\times 10^{32}, 5.4\times 10^{32}$ and $\rm 5.0\times
10^{31}$\,W\,MHz$^{-1}$ for the OH lines at 1612 and 1665/1667\,MHz
and the H$_2$O line, respectively. These are shown in Fig.~3 alongside
observations of masers in other IR-luminous galaxies.

\begin{figure}
\centerline{\psfig{file=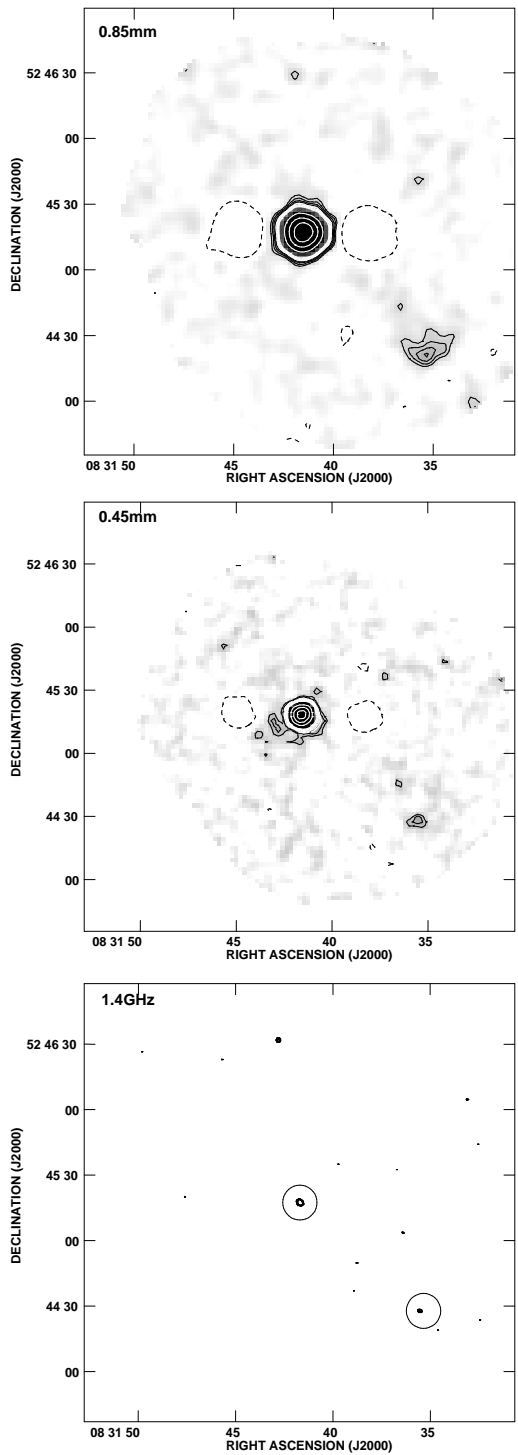,angle=0,width=2.7in}}
\vspace{-0.4cm}
\noindent{\small\addtolength{\baselineskip}{-3pt}}
\caption{Greyscales of the APM\,08279+5255 field, as imaged by SCUBA
at 850\,$\mu$m (top) and 450\,$\mu$m (middle) and by the VLA at
1,400\,MHz (lower). In the submm, contours are plotted at
$-$3,3,4...10,20...100 $\times\sigma$.  These are signal-to-noise
images, smoothed to 15 and 7.5\,arcsec {\sc fwhm}.  The new SMG,
SMM\,J083135.5+524425, can be seen to the south west of
APM\,08279+5255, the brightest object. Their $-$0.5/+1.0/$-$0.5 beam
profiles are characteristic of real sources in uncleaned jiggle
maps. In the radio, contours are plotted at $-$4,4...10 $\times\sigma$
where $\sigma$ is 28\,$\mu$Jy\,beam$^{-1}$ and the synthesised beam
measures $\rm 1.45\times 1.16$\,arcsec {\sc fwhm} with its major axis
at position angle 52$^{\circ}$. Circles (8-arcsec radius) mark the
850-$\mu$m positions of APM\,08279+5255 and the new SMG.}
\end{figure}

\section{Ancilliary data: submm, radio and IR}

The APM\,08279+5255 field was imaged with the SCUBA bolometer array
(Holland et al.\ 1999) during 2000 March to 2001 March, primarily to
help characterise the point spread function for the deep submm survey
of high-redshift radio galaxies (M00AU09) presented by Stevens et al.\
(2003).

The James Clerk Maxwell Telescope's (JCMT's) secondary mirror was
chopped and nodded by 30\,arcsec in R.A., resulting in the
characteristic $-$0.5/+1.0/$-$0.5 beam profile visible in
Fig.~4. After overheads, 25.6$\,$ks were spent integrating on the
field, split into approximately hour-long chunks separated by checks
on focus, pointing accuracy and atmospheric opacity. The data were
reduced using standard software, but were imaged using several new
techniques. First, the high flux density of APM\,08279+5255 allowed us
to accurately align each 2,560-s of data, peaking up at 450\,$\mu$m
having adopted the 1,400-MHz radio position for APM\,08279+5255 (see
later; R.A.\ $\rm 08^h 31^m 41^s.708$, Dec.\ $\rm +52^{\circ} 45'
17.44''$ J2000). Despite the small shifts being applied (1.6 and
0.8\,arcsec r.m.s.\ in R.A.\ and Dec., consistent with normal pointing
drifts), this `shift-and-add' approach made an appreciable difference
to the final image. Second, we employed new versions of the {\sc surf}
{\tt setbolwt} and {\tt rebin} tasks, developed for the radio galaxy
survey. These enabled us to determine accurately weighted and
calibrated signal and noise values for square pixels of arbitrary
size, independently of neighbouring pixels, at 13-arcsec {\sc fwhm}
resolution for 850\,$\mu$m.  The signal and noise for each pixel
correspond to the stream of data collected only when bolometers are
centred within that pixel in a manner reminiscent of the
`zero-footprint' approach used by Scott et al.\ (2002). For Fig.~4 we
smoothed with 6- and 3-arcsec {\sc fwhm} Gaussians and the resulting
850- and 450-$\mu$m maps have noise levels of 1.7 and
7\,mJy\,beam$^{-1}$ (before cleaning).

This led to the detection of an new, relatively bright SMG in the
vicinity of APM\,08279+5255, with R.A.\ $\rm 08^h 31^m 35^s.46$
(35$^s$.48), Dec.\ $\rm +52^{\circ} 44' 22.4''$ (26.8$''$) J2000 at
850\,$\mu$m (450\,$\mu$m), with a positional uncertainty of
2.2\,arcsec at 850\,$\mu$m (1\,$\sigma$, Ivison et al.\ 2005). We
denote this source as SMM\,J083135.5+524425, adopting the mean submm
position and following the IAU convention. Using the quasar to align
the 450- and 850-$\mu$m coordinate frames would result in a marginal
(0.9-arcsec) reduction of the 4.4-arcsec offset between the 450- and
850-$\mu$m positions.  The peak 850-$\mu$m flux density of the new
source, bootstrapping the calibration from Uranus (via
APM\,08279+5255: $\rm 286\pm 10$ and $\rm 82.7\pm 2.2$\,mJy at 450 and
850\,$\mu$m; cf.\ 84 and 285\,mJy, Barvainis \& Ivison 2002), is $\rm
11.2\pm 1.7$\,mJy, making it a relatively bright SMG (we would expect
around one per 100\,arcmin$^2$). The flux density rises only
marginally, to 11.4\,mJy, when measured in an 14-arcsec-radius
aperture, so any evidence that the source is extended is marginal at
best. The SMG is also detected at 450\,$\mu$m with a flux density of
$\rm 35\pm 7$\,mJy, though only following our shift-and-add
treatment. The 450-/850-$\mu$m flux ratios of the new SMG and
APM\,08279+5255 are consistent ($\sim$8) to within the large
photometric uncertainties.

APM\,08279+5255 was then targeted at 1,400\,MHz in an attempt to
further constrain the position and redshift of the new SMG. For this
we used the VLA in its A configuration, with correlator mode 4 (28
$\times$ 3.125-MHz channels, half each for left and right circular
polarisations), during 2001 January. A noise level of
28\,$\mu$Jy\,beam$^{-1}$ was achieved. SMM\,J083135.5+524425 has a
single, bright and relatively compact radio counterpart centred at
R.A.\ $\rm 08^h 31^m 35^s.59$, Dec.\ $\rm +52^{\circ} 44' 27.8''$
J2000 ($\sim$0.2-arcsec uncertainity, 1\,$\sigma$), with a total flux
density of $\rm 386\pm 55$\,$\mu$Jy (after correcting for bandwidth
smearing and the primary beam response).

The total 1,400-MHz flux density of APM\,08279+5255 was measured to be
$\rm 3.05\pm 0.07$\,mJy so, as with the submm flux ratios discussed
earlier, the submm/radio flux ratios ($\sim$28) of the new SMG and
APM\,08279+5255 are consistent with one another to within the
photometric uncertainties. This suggests they may well lie at a common
redshift. A full track with the IRAM Plateau de Bure Interferometer,
tuned to CO(4--3) at $z=\rm 3.911$, would confirm or refute this
possibility. Although perfectly consistent with a point source, we
note that the morphology of the new SMG is suggestive of an arc and so
we cannot rule out the possibility that this is a new lens feature
caused by an unknown, unseen, massive foreground cluster, though we
would then expect to see arcs from the more numerous optical/IR
population.

We have inspected a $K_{\rm s}$-band image, obtained by us during 2000
November 12 in 1.0-arcsec seeing using the INGRID camera on the 4.2-m
William Herschel Telescope (WHT), at the position of the radio
emission. The region is blank in the near-IR, with
$K_{\rm s}>\rm 20.5$ (3$\sigma$, 1.5-arcsec radius aperture,
Vega). Fig.~4 also reveals faint (3$\sigma$), coincident 450- and
850-$\mu$m emission, $\sim$15\,arcsec north east of
SMM\,J083135.5+524425, this time with no radio or IR counterpart.

\section{Concluding remarks}

We have conducted a preliminary search for OH and H$_2$O gigamaser
emission from the distant, luminous quasar, APM\,08279+5255. Our most
pessimistic prediction -- an OH maser exhibiting behaviour of the form
$\loh \propto \lfir^{1.2}$, subject to extreme lens amplification --
would have been within range of our search. We have thus set a
meaningful constraint on the OH maser luminosity of APM\,08279+5255,
though we acknowledge that FIR luminosity is not the only criterion
for OH masing.

We note that 10$\times$ more bandwidth is already available at the
GMRT than was used here. Upcoming facilities will offer further
improvements and we suggest that the Square Kilometre
Array\footnote{http://www.skatelescope.org/}, for example, should be
designed to have sufficient instantaneous bandwidth to employ this
technique effectively on blank-field SMGs of the type expected to be
found in their thousands in the coming few years.

Our submm imaging has uncovered a new, bright SMG $\sim$75\,arcsec
from the quasar, possibly within a physical distance of $<$1\,Mpc
since their submm/radio flux ratios are a good match. Its position has
been determined accurately in the radio, but it has no IR counterpart.

The effectiveness of the shift-and-add technique suggests deep
archival searches could be made in the vicinity of other bright
sources, e.g.\ JCMT pointing sources such as 3C\,345 and BL\,Lac that
have accumulated many hours of integration. Blank-field surveys with
the next generation of sensitive submm cameras may also benefit from
variants of this technique, particularly at 450\,$\mu$m.

\section*{Acknowledgements}

Many thanks to Andy Biggs, Chris Carilli, Thomas Greve, Geraint Lewis,
Nuria Lorente, Ian Smail, Lister Stavely-Smith, Jason Stevens, Richard
Townsend and to an anonymous referee for some astute advice.  The
National Radio Astronomy Observatory is operated by Associated
Universities Inc., under a cooperative agreement with the National
Science Foundation. The JCMT is operated by the Joint Astronomy Centre
in Hilo, Hawaii, on behalf of the Particle Physics and Astronomy
Research Council in the UK, the National Research Council of Canada,
and The Netherlands Organisation for Scientific Research. The WHT is
operated on the island of La Palma by the Isaac Newton Group in the
Spanish Observatorio del Roque de los Muchachos of the Instituto de
Astrofisica de Canarias.

\end{document}